\documentclass[12pt,preprint]{aastex}
\usepackage{times}
\usepackage{natbib}
\slugcomment{Astrophysical Journal Letters: Sunrise Special Issue,\\ 
             reveived 2010 June 16; accepted 2010 August 2}

\shorttitle{Vortex tubes in solar granulation}
\shortauthors{O. Steiner et al.}

\begin{document}


\title{Detection of vortex tubes in solar granulation from observations with {\sc Sunrise}}


\author{\textsc{O.~Steiner$^1$, M.~Franz$^1$, N.~Bello Gonz{\'a}lez$^1$, Ch.~Nutto$^1$, 
R.~Rezaei$^1$, V.~Mart{\'{\i}}nez Pillet$^2$, J.~A.~Bonet Navarro$^2$, J.~C.~del~Toro~Iniesta$^3$, V.~Domingo$^4$, S.~K.~Solanki$^{5,6}$, M.~Kn{\"o}lker$^7$, W.~Schmidt$^1$, 
P.~Barthol$^5$, and A.~Gandorfer$^5$}}

\affil{$^1$Kiepenheuer-Institut f\"ur Sonnenphysik, Sch\"oneckstrasse 6,
       79104 Freiburg, Germany}
\affil{$^2$Instituto de Astrof{\'{\i}}sica de Canarias, 38200, La Laguna, Tenerife, Spain}
\affil{$^3$Instituto de Astrof\'{\i}sica de Andaluc\'{\i}a (CSIC), Apdo.~de Correos 3004, 
       18080 Granada, Spain}
\affil{$^4$Grupo de Astronom\'{\i}a y Ciencias del Espacio, Universidad de Valencia, 
       46980 Paterna, Valencia, Spain}
\affil{$^5$Max-Planck-Institut f\"ur Sonnensystemforschung, 37191 Katlenburg-Lindau, Germany}
\affil{$^6$School of Space Research, Kyung Hee University, Yonging, Gyeonggi 446-701, Republic of Korea}
\affil{$^7$High Altitude Observatory, National Center for Atmospheric Research (NCAR),
      Boulder, CO 80307-3000, USA}

\email{steiner@kis.uni-freiburg.de}

\begin{abstract}
We have investigated a time series of continuum intensity maps and corresponding Dopplergrams
of granulation in a very quiet solar region at the disk center, recorded with the Imaging 
Magnetograph eXperiment (IMaX) on board the balloon-borne solar observatory 
{\sc Sunrise}. We find that granules frequently show substructure in the form of 
lanes composed of a leading bright rim and a trailing dark edge, 
which move together from the boundary of a granule into the granule itself. We find 
strikingly similar events in synthesized intensity maps from an ab initio numerical 
simulation of solar surface convection. From cross sections through the computational 
domain of the simulation, we conclude that these `granular lanes' are the visible signature 
of (horizontally oriented) vortex tubes. The characteristic optical appearance of 
vortex tubes at the solar surface is explained. 
We propose that the observed vortex tubes 
may represent only the large-scale end of a hierarchy of vortex tubes existing near the 
solar surface.
\end{abstract}



\section{Introduction}
\label{sec:intro}

Solar granulation is the superficial, visible manifestation of the motion of plasma
in the convection zone of the Sun. Warm plasma ascends to the solar surface within 
granules, cools radiatively at the surface, and returns back into the convection zone 
within the intergranular lanes. The difference in temperature between 
the warm ascending and the cool descending plasma accounts for the observed intensity 
contrast between the bright granules and the network of dark intergranular lanes
and for the convective blueshift, the limb effect, and the C-shape of photospheric 
spectral lines \citep{stix2002}.
For a good comprehension of the granular phenomenon, it has proven of value to think of 
the surface cooling and the subsequent descent of cool plasma as the primary process 
that drives convection near the surface 
\citep{spruit1997,rast1999,dmueller+al2001}. 

Thus, the concept of an ``average granule'' was developed with a mean velocity field
and mean values for lifetime and effective diameter 
\citep[see][and references therein]{stix2002}.
Although fundamentally correct, this picture of quasi-laminar flow cannot 
account for the fine structure of granulation, which has become 
apparent through
observations with the balloon-borne solar observatory {\sc Sunrise} 
\citep{barthol+al2010_shortt,solanki+al2010_shortt}. 
Frequently occurring fine structure elements are `granular lanes', typically consisting 
of a bright and a dark edge, which travel from the visible boundary of granules into the 
granule itself. In this Letter, we present several events in maps of the 
continuum intensity and the corresponding Dopplergram (Sections~\ref{sec:obs} and 
\ref{sec:fine}). We reveal their physical nature with the help of numerical 
simulations as the visible signature of vortex tubes (Sections~\ref{sec:sim} and
\ref{sec:vt}) and draw conclusions in Section~\ref{sec:conclusions}.


\section{Observation and data reduction}
\label{sec:obs}

The observational data were obtained on 2009 June 9
between 00:35:49~UT and 02:02:48~UT with the Imaging Magnetograph eXperiment (IMaX), 
on board {\sc Sunrise}.  
For a detailed description of the IMaX instrument and the data reduction procedure, we 
refer to \citet{martinez-pillet+al2010_shortt}. 
Images were taken for four polarimetric states with the etalon being centered at the 
wavelength positions $\pm 80$, $\pm40$, and $+227$~m\AA\ with respect to the core of 
the spectral line \ion{Fe}{1} at 5250.2~\AA\ ($g = 3$, $\chi = 0.121$~eV). 
The continuum images have an effective exposure time of 6~s.
All images were reconstructed using deconvolution with a point-spread function (PSF)
and Wiener-Helstrom filtering. The PSF was obtained from phase diversity image pairs 
taken close in time to the measurements. 
We obtained two consecutive series of science images that together cover a time period of
53.9~minutes with a cadence of 33~s. The noise level is $10^{-3}$ of the continuum
intensity for the non-reconstructed and  (2.5--3)$\times 10^{-3}$ for the reconstructed data.
A spatial resolution of 0.15\arcsec--0.18\arcsec\ in all Stokes parameters was achieved 
by combining the onboard image stabilization system \citep{berkefeld+al2010_shortt} 
with the phase-diversity-based post-processing technique. 
These data show granulation of a very quiet solar region in a field of view (FOV) of 
approximately $51.5\arcsec \times 51.5\arcsec$ at the disk center. 
The Doppler velocity, $v_{\rm{D}}$, is inferred from a Gaussian fit to the
intensities at the four wavelength points in the spectral line, after 
normalization to the local continuum. In addition, a spatially dependent 
blueshift over the FOV 
was taken into account in the course of the fit procedure.
The zero point of the velocity corresponds to the mean wavelength 
position over the full FOV, corrected for a convective blueshift of 
0.2~km\,s$^{-1}$. 
We estimate the absolute error of $v_{\rm{D}}$ to
about $\pm 0.2$~km\,s$^{-1}$.


\section{Fine structure elements of granules}
\label{sec:fine}

\begin{figure}
\epsscale{1.0}
\plotone{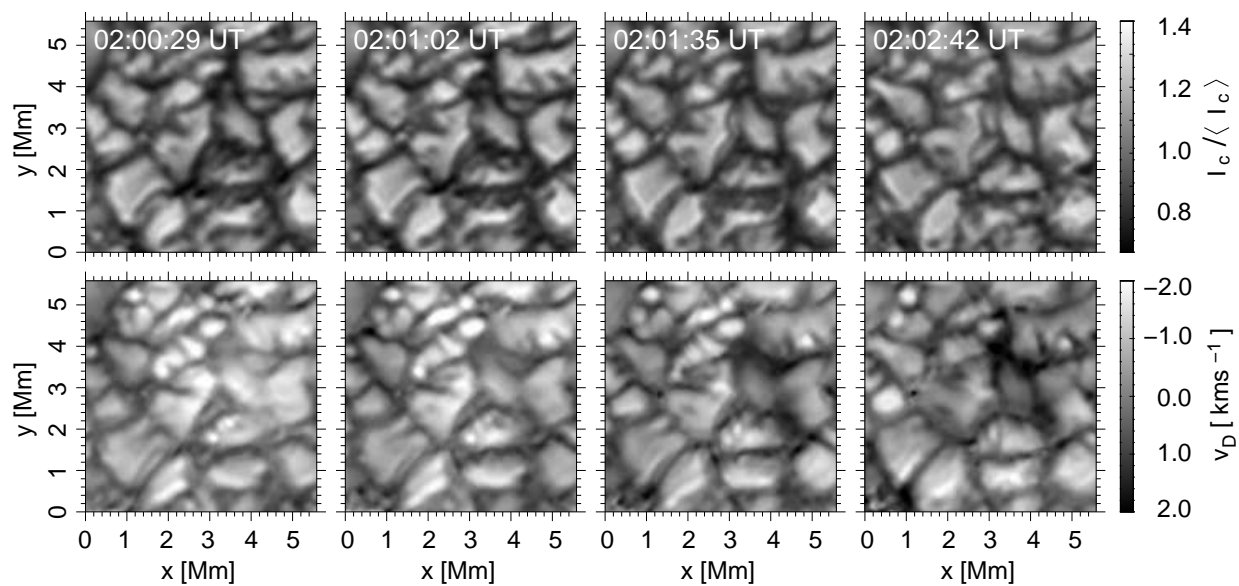}
\caption{Time series showing the development of a bright/dark granular lane at the 
right boundary of the granule in the image center at $(x,y)=(2.5,2.5)$~Mm. Top row: 
sequence of reconstructed images in the continuum intensity at $\lambda=5250.4$~\AA. 
Bottom row: corresponding 
sequence of Doppler maps. The maps are small subfields of the full FOV, 
measuring $5.6\times 5.6$~Mm each. Positive velocities correspond to downdrafts.
\label{fig:tseqobs}}
\end{figure}

Figure~\ref{fig:tseqobs} shows the time sequence of the formation and evolution of 
the type of granular fine structure, which we consider in the present investigation. 
The top row shows the reconstructed intensity maps. 
From this sequence one can see the following:
the granule in the center of the images ($(x,y) = (2.5,2.5)$~Mm) expands rapidly,
apparently colliding with the two neighboring granules on the right hand side. At time 
$t={\textrm{02:00:29 UT}}$, the right border of the expanding granule starts to 
become slightly enhanced in intensity at $(x,y) = (2.8,2.7)$~Mm. 
In the next frame---33~s later---a bright rim detaches from the border of the granule
and moves into the granule itself. The ends of the 
bright lane, however, remain attached to the granule border. 
At $t={\textrm{02:01:35 UT}}$, a second rim, similar in brightness, develops in the 
upper right side of the granule, connecting to the existing one at 
$(x,y) = (2.8,2.9)$~Mm, and complementing it to a Y-shaped bright lane.
In the course of this development, a dark lane immediately adjacent but trailing 
the bright lane appears, giving the impression of a sharp edge to the trailing
side of the bright lane. This sequence of events is visible in both the 
non-reconstructed (not shown here)
and the reconstructed images and is therefore not an artifact of the reconstruction.
The observing sequence stops at 02:02:42~UT. Typically, such bright/dark 
`granular lanes' either retreat back to the boundary of the granule or disappear 
with the subsequent dissolution of the granule. They should not be mistaken with
the dark intergranular lanes.

The sharp separation of the right boundary region from the 
rest of the granule by the granular lane  
is less pronounced in the Doppler maps (the bottom row of Figure~\ref{fig:tseqobs}), 
where the granule appears undivided and still intact. The granular lane is visible
only at the beginning of the sequence but is barely noticeable at times 
later than $t=\textrm{02:02:09 UT}$. The bright lane and the area to the left (ahead) 
of it show flows in the upward direction with speeds up to $-1.8$~km\,s$^{-1}$. 
Most of the area behind the bright rim, including the dark lane, harbors upflows as well.
Close to the border of the granule and within the adjacent intergranular lane,
the plasma flows in the downward direction with speeds up to 1.2~km\,s$^{-1}$.

\begin{figure}
\epsscale{1.0}
\plotone{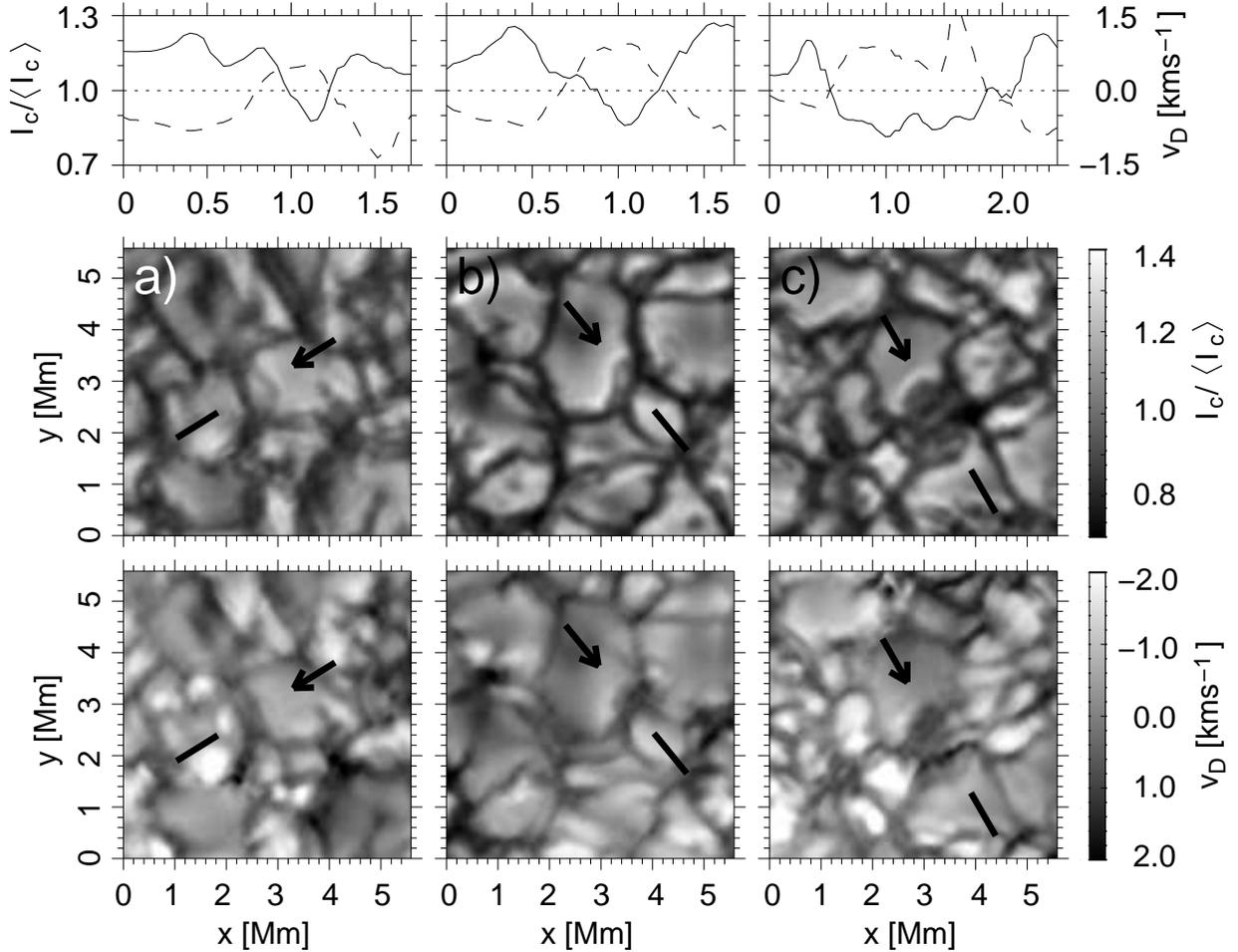}
\caption{Instances of granular lanes. Middle row: 
reconstructed images of the continuum intensity at $\lambda=5250.4$~\AA. 
Bottom row: corresponding sequence of Doppler maps. Top row: plots
of the relative intensity (solid curve) and the Doppler velocity in km\,s$^{-1}$ 
(dashed curve) in a section running from the
tip of the arrow to the opposite inner end of the black mark, indicated in 
the intensity and Doppler maps. Positive velocities indicate downdrafts,
distances are given in Mm.
\label{fig:eventobs1}}
\end{figure}

\begin{figure}
\epsscale{1.0}
\plotone{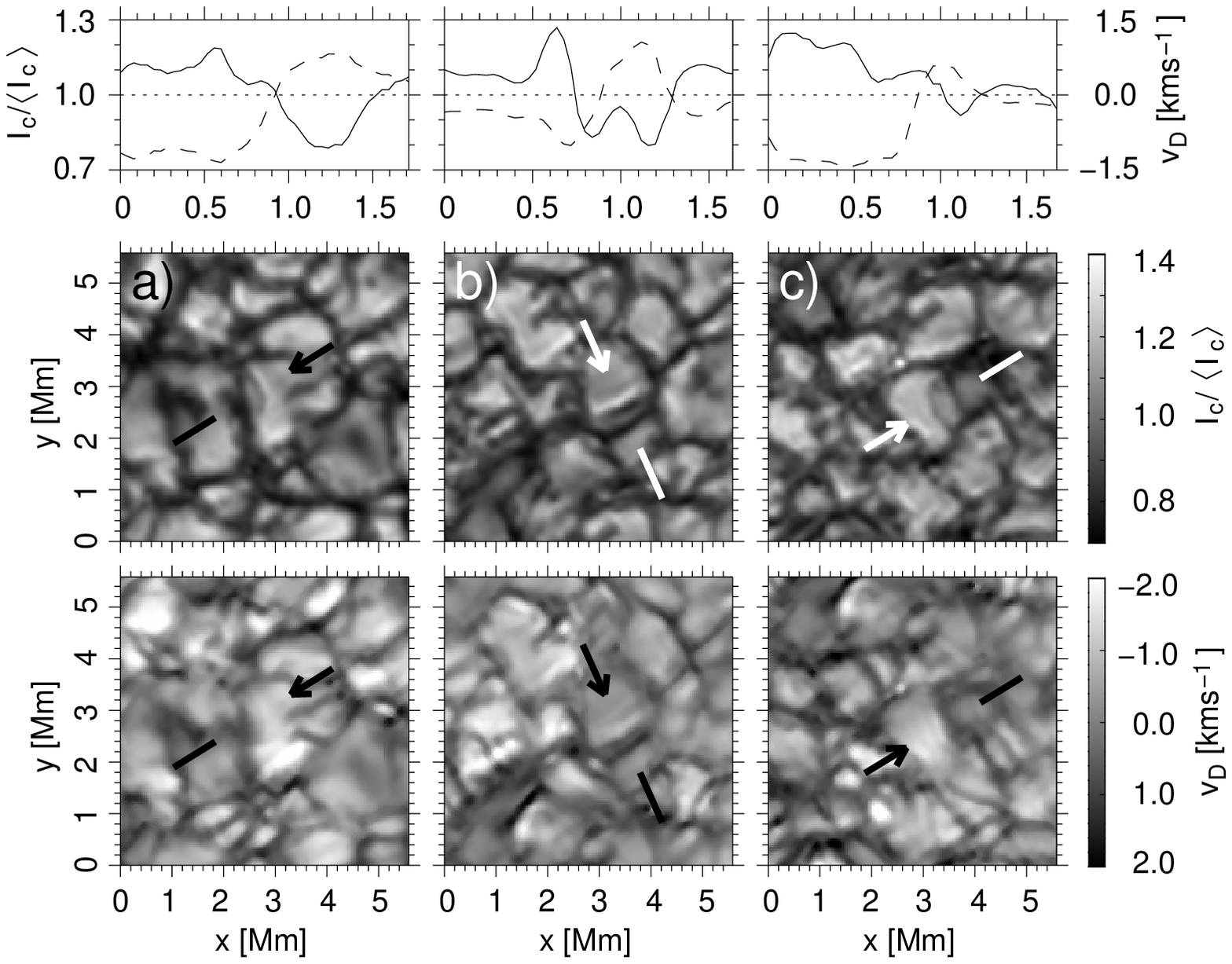}
\caption{Instances of granular lanes depicted as in Figure~\ref{fig:eventobs1}.
\label{fig:eventobs2}}
\end{figure}

Figures~\ref{fig:eventobs1} and \ref{fig:eventobs2} show six instances of 
events similar to the one shown in 
Figure~\ref{fig:tseqobs}. The middle row shows reconstructed intensity maps; located
below are the corresponding Doppler maps. The panels in the top row show plots of the
intensity and the Doppler velocity along the cuts indicated in the maps by the black
arrows and marks. The abscissa of each plot runs from the tip of the arrow
to the inner end of the opposite mark. 

Typically, granular lanes have an oval outline (cf.~Figures~\ref{fig:eventobs1}(a)-(c)),
but straight lanes occur as well (cf.~Figures~\ref{fig:eventobs2}(b) and (c)). 
Some bright lanes lack their distinct dark counterpart, like in Figure~\ref{fig:eventobs1}(c), 
and sometimes there exists a solitary dark lane or a darkish area moving into the 
granule without a distinct leading bright lane (cf.~Figure~\ref{fig:eventobs2}(c)). 
Sometimes the trailing dark lane starts to separate from the leading bright lane,
lagging behind, as is apparent in the last image of the time sequence
of Figure~\ref{fig:tseqobs}.

Flows are always in the upward direction within and ahead of the bright 
lane, with typical speeds of $-0.8$ to $-1.4$~km\,s$^{-1}$. Also within the trailing 
dark lane, there are mostly updrafts of similar speeds. 
The area behind the bright/dark lane shows upflows and/or downflows. The adjoining
intergranular lane always shows downdrafts with typical speeds of 1.0 km\,s$^{-1}$. 
The edge of the bright/dark lanes moves in the horizontal direction with proper 
velocities between approximately 1 and 4 km\,s$^{-1}$. The observed lanes 
show lifetimes from 1.5 to 6.5~minutes. 
Within a time period of 53.9~minutes and the given FOV, we found 21 events of granular lanes,
i.e., 0.017 events Mm$^{-2}$\,hr$^{-1}$.
This number must be considered a lower limit since we picked only the 
most salient events.


\section{Numerical simulation}
\label{sec:sim}

The simulation was carried out with the 
CO\raisebox{0.5ex}{\footnotesize 5}BOLD code \citep{freytag+al2002}. It
solves the coupled system of the equations of compressible magnetohydrodynamics 
in an external gravity field and with non-local, frequency-dependent radiative transfer 
in three spatial dimensions. Further details of the method can be found in 
\citet{schaffenberger+al2005}. 

\begin{figure}
\epsscale{1.0}
\plotone{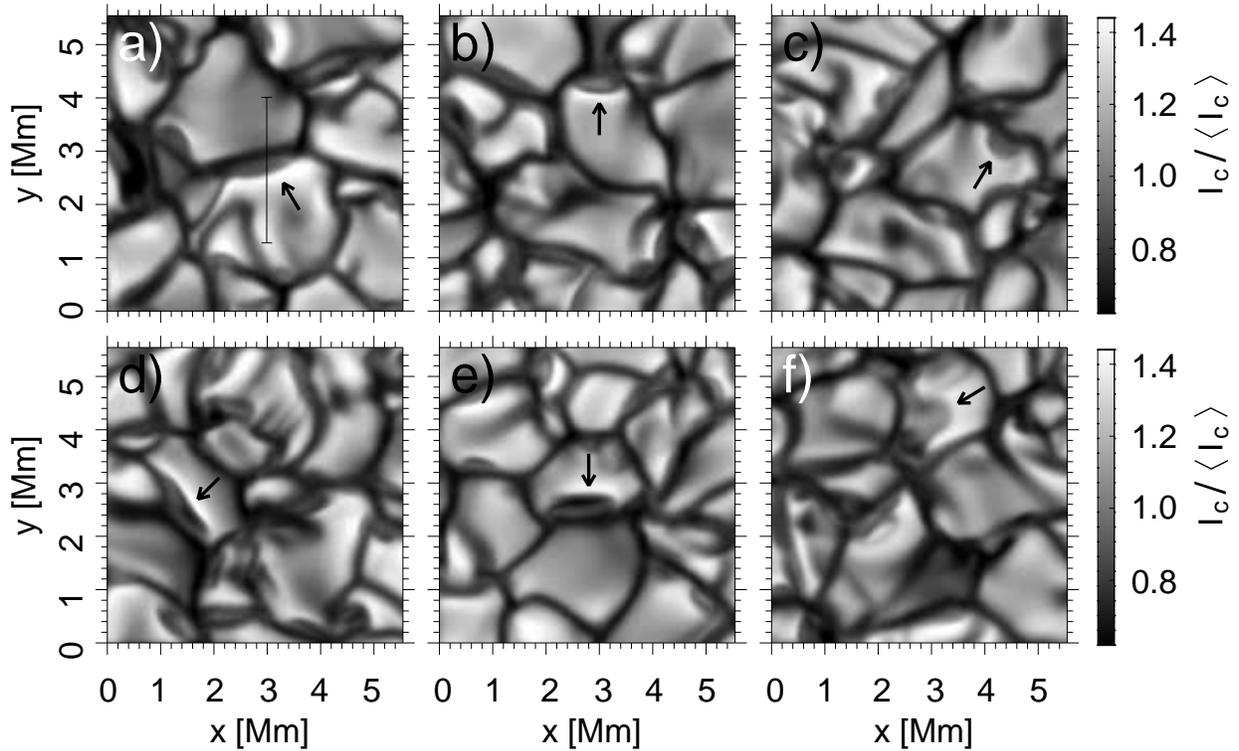}
\caption{Instances of granular lanes (marked by arrows) from the numerical simulation. 
The images show the synthesized continuum intensity at 5250~\AA\ in a subfield
of 5.6~Mm $\times$ 5.6~Mm of the full simulation domain. The black line
in panel (a) indicates the location and extension of the vertical cross section of 
Figure~\ref{fig:vtsection}. 
\label{fig:eventsim}}
\end{figure}

The computational domain extends over a height range 
of 2.8\,Mm in which the mean surface of optical depth 
unity is located in the middle. The horizontal extension is 9.8\,Mm.  
The spatial resolution is 40\,km in the horizontal direction, while it increases
continuously in the vertical direction from 50~km in the lower part to 20~km
in the middle and upper part. The simulation domain contains a weak magnetic field,
which was advected by rising plumes through the bottom boundary, similar as in
\citet{steiner+al2008}.
The simulation was carried out for a real time period of 55.7~minutes from which
the emergent continuum intensity at 6300~\AA\ was synthesized with a cadence of 
5~s. For the synthetic intensity maps of Figure~\ref{fig:eventsim}, we have computed
the intensity at 5250~\AA\ for a better comparison with the IMaX data.

Figure~\ref{fig:eventsim} shows subfields of synthesized intensity maps. 
Figure~\ref{fig:eventsim}(a) shows a pair of
granular lanes resembling the event shown in Figure~\ref{fig:tseqobs}. 
Other maps show events strikingly similar to the ones of Figures~\ref{fig:eventobs1}
and \ref{fig:eventobs2}, 
including a bright lane without trailing dark lane (Figure~\ref{fig:eventsim}(c)), an 
exceptionally dark, straight lane (Figure~\ref{fig:eventsim}(e)), and a darkish area 
moving into the granule lacking a leading bright/dark lane (Figure~\ref{fig:eventsim}(f)).

We find 32 events, i.e., 0.37 events Mm$^{-2}$\,hr$^{-1}$, which
is more frequent than in the observations.
Similar to the observations, the lanes move horizontally with speeds of 
1--5~km\,s$^{-1}$, have a lifetime of several minutes, seem to originate from 
collisions with neighboring granules, and sometimes retreat back to the
boundary of the granule or dissolve together with the granule. The simulations
show a larger variety of appearances of granular lanes than the observations
do, including solitary dark lanes, ring like lanes, or the passage of a granular lane
across an intergranular lane to the neighboring granule.


\section{Identification with vortex tubes}
\label{sec:vt}

\begin{figure}
\epsscale{1.0}
\plotone{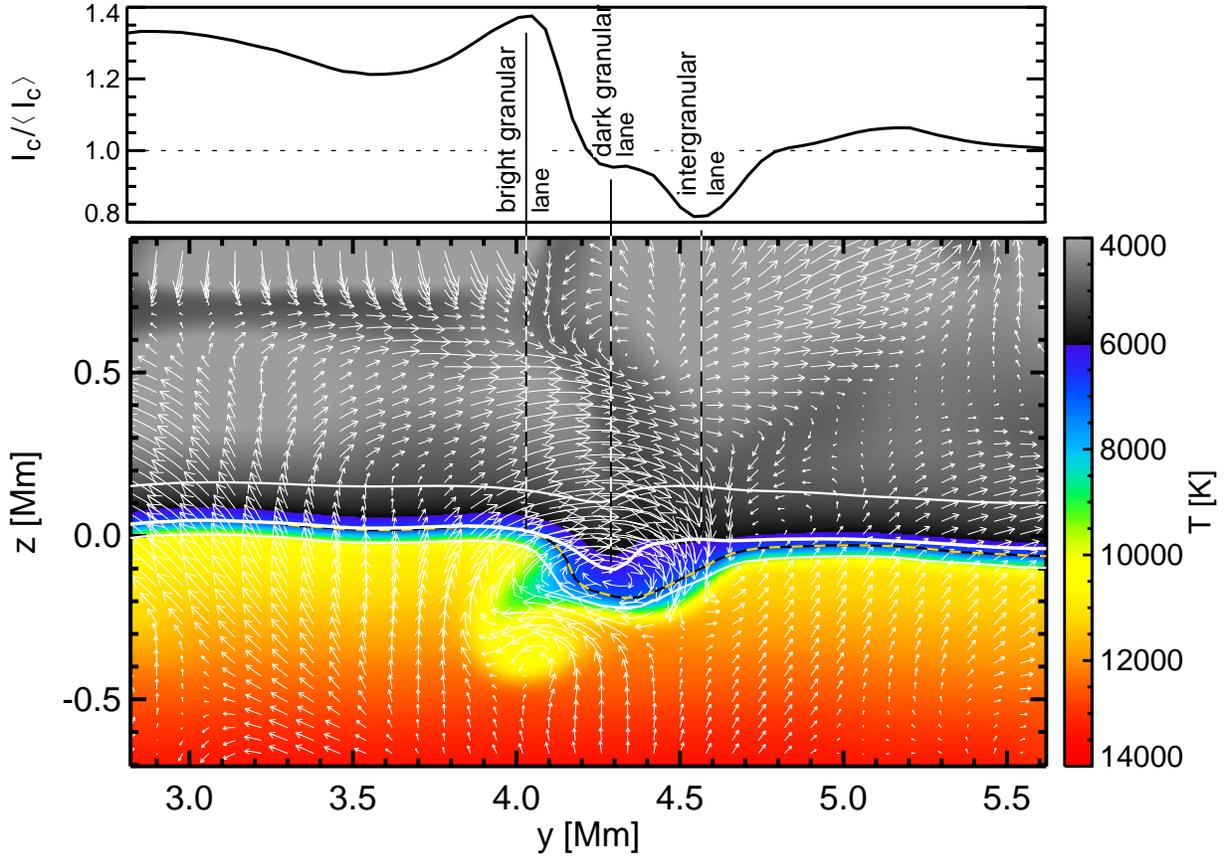}
\caption{Vertical cross section through the computational domain for the event and 
at the location shown in Figure~\ref{fig:eventsim}(a). The top panel shows the emergent
synthesized continuum intensity at $\lambda=5250$~\AA. Colors indicate temperature and
the arrows indicate the velocity field. The white contours indicate optical depths of
$\log\tau_{\rm c}=-1$, 0, and 1, from top to bottom, respectively, the yellow/black 
dashed contour represents the 7000~K isotherm. The longest arrows correspond to a 
velocity of 8.5~km\,s$^{-1}$.
\label{fig:vtsection}}
\end{figure}

In order to reveal the physical nature of these objects, we now study
a vertical cross section through the computational domain at the location of a
representative granular lane. Figure~\ref{fig:vtsection} is a cross section through
the granular lane of Figure~\ref{fig:eventsim}(a), at the position and of length as
indicated by the black line in Figure~\ref{fig:eventsim}(a). It shows the temperature 
(color scale),
together with the velocity projected on the plane of section (arrows), and
the continuum optical depths of $\log\tau_{\rm c}=-1.0$, 0.0, and 1.0 (white, 
horizontally running contours). The yellow/black dashed contour refers to the 7000~K 
isotherm. The panel on top of the cross section shows the corresponding synthesized 
continuum intensity at $\lambda=5250$~\AA.

From Figure~\ref{fig:vtsection}, it becomes immediately evident that the granular 
lane is caused by the vortical flow located at $(y,z) = (4.3,-0,1)$~Mm.
For all the instances of granular lanes shown in Figure~\ref{fig:eventsim}, 
and for additional 20 events, we have verified that they were all due to 
vortical flows similar to the one shown in Figure~\ref{fig:vtsection}.
The vorticities of these flows have a horizontal orientation and show a
coherent structure over a granular scale. In the following we shall refer to 
these objects as ``vortex tubes'' following turbulence theory.

A common property of all the vortex tubes that we have inspected is that  
they contain relatively cool material. As a consequence, the surface of optical 
depth unity is always significantly depressed at the location of the vortex tube, 
while the maximum depression coincides approximately with the location of the tube 
axis, as is the case in Figure~\ref{fig:vtsection}. Warm plasma flowing upward must 
circumvent this cool material, which presents an obstacle to the general heat flow. 
The deviated warm material joins the upflowing side of the vortex tube, appearing 
at the surface where the 7000~K isotherm in Figure~\ref{fig:vtsection} approaches or 
even surpasses the surface of $\tau_{\rm c} = 1$. This configuration
causes the bright lane at $y=4.05$~Mm. Starting from this location, the isotherm 
drops rapidly toward the vortex tube---more steeply than the 
$\tau_{\rm c} = 1$ surface---and causes the sharp edge between the bright lane 
and the adjoining dark plateau. In the particular case investigated here, no 
pronounced dark lane exists because the area further downstream of the vortex-tube 
axis stays dark (also cf.~Figure~\ref{fig:eventsim}(a)) and does not show an increase 
in intensity as in other cases. Upstream and within the bright lane, 
the velocity has a significant component in the upward direction in agreement
with the observations. In the range of the dark area and immediately above the 
vortex tube, plasma flows in a rather horizontal direction. This flow is expanding,
accelerated by the increasing horizontal gas-pressure gradient between the 
interior of the granule and the region above the vortex tube. Directly above the 
vortex tube, the flow assumes transonic speeds and gas pressure and temperature 
have a local minimum. Since pressure and temperature are low, the opacity above the 
vortex tube is reduced (dip in the $\log \tau = -1$ surface), which opens a relatively 
transparent view to the cool interior of the vortex tube. This is the origin of 
the dark lane, which trails the bright granular lane. Further downstream, the horizontal
wind decelerates and collides with flows from the neighboring granule. In some cases
the intensity in this region increases again (but not in the case of 
Figure~\ref{fig:vtsection}) before the flow turns into the downdraft, 
which causes the dark intergranular lane at $x=4.6$~Mm. 

In most cases, the central axis of the vortex tube is relatively steady,
staying close to but below the surface of $\tau_{\rm c}=1$.
It is rather the shape and size of the vortex tube, which changes in time 
and causes the horizontal displacement of the bright lane and the expansion 
of the trailing darkish area. Sometimes vortex tubes start with an oblique 
cross section, with the major axis being oriented in the vertical direction and
evolve into an oblique, flat cross section with major axis  oriented in the 
horizontal direction. In particular, extended dark areas moving into granules 
(cf.~Figures~\ref{fig:eventobs2}(c) and \ref{fig:eventsim}(f)) are caused by 
strongly flattened vortex tubes, which start to resemble vortex sheets.
It is remarkable that the downdraft side of the vortex tube hardly ever 
shows a deep downdraft. Rather, this downdraft is shallow and runs into the
vortex flow, where it gets recycled and contributes to the growth of the
vortex tube. However, this state of matter is hidden from the visible
surface so that in the intensity map the location of the downdraft coincides 
with a normal intergranular lane (cf.~Figure~\ref{fig:eventsim}(a)). 
Without having carried out a systematic study yet, it seems that an initial, 
regular intergranular downdraft may get wrapped into a vortex tube in the 
course of a collision of two neighboring granules. This would explain the 
absence of deep downdrafts at the location of vortex tubes. This
scenario is confirmed by the fact that usually a regular downdraft is
present outside of both ends of the vortex tube.

The cores of closed stream lines in the vortex tubes have a typical mean
radius of 150~km. Velocities along a stream line at this distance 
are typically 3~km\,s$^{-1}$ on the lateral sides, 8~km\,s$^{-1}$ on the
top side, and 5~km\,s$^{-1}$ on the bottom side. Thus, one 
revolution takes about half a minute, which is considerably shorter than the 
lifetime of typical vortex tubes.

Sometimes, the bright lane is missing, e.g., when the upflow has enough 
lateral space to escape away from the vortex tube in the horizontal direction. 
In this case a solitary dark lane forms above the vortex-tube center. 
Sometimes a flat vortex tube stays at the surface of a granule having the 
appearance of a normal intergranular lanes, which, however, does not harbor 
a deep downdraft. Conspicuously bright, small granules are often the result 
of two neighboring vortex tubes with opposite sense of rotation.

In the simulation, we see occasional magnetic field intensification at locations 
of vortex tubes---a few cases rather show field expulsion. Whenever magnetic fields 
are present, they tend to point in the horizontal direction near or above the 
$\tau_{\rm c}=1$ surface. Hence, vortex tubes might be a significant source of 
horizontal magnetic fields in the photosphere. 
Patches of horizontal fields are found to be preferentially located near the boundaries
of granules \citep{ishikawa+al2010,danilovic+al2010_shortt}, which coincides with the 
location of vortex tubes. 
However, our observations, which include the polarization signal, do not 
indicate enhanced magnetic fields at locations of granular lanes.


\section{Conclusions}
\label{sec:conclusions}

From observations with the imaging magnetograph IMaX on board the
balloon-borne solar observatory {\sc Sunrise}, we find granular fine structure, 
consisting of moving lanes, which are usually composed of a leading bright rim and a 
trailing dark edge. They move from the border of a granule into
the granule itself. Sometimes they retreat back again before they disappear.
Although we have noticed these features for the first time in Sunrise data, 
after completing the analysis we also looked at other data sets and found them 
there as well.

With the help of a numerical simulation of solar surface convection, we identify 
these objects as vortex tubes. 

From the equation of vorticity {\boldmath$\omega$} for an
inviscid, compressible medium,
\begin{equation}
\label{eq:vorticity}
\frac{{\rm D}\mbox{\boldmath$\omega$}}{{\rm D} t}  = 
\nabla (\mbox{\boldmath$v$}\cdot\mbox{\boldmath$\omega$})
+ \frac{1}{\rho^2}\nabla\rho\times\nabla p\;,
\end{equation}
one can see that the baroclinic term (second term on the right hand side)
can be a source of vorticity. This term grows important for non-isentropic  
flows, which indeed exist at the solar surface, where radiative transfer
takes place. In particular, vorticity is generated near the boundaries of
granules and in the intergranular downdrafts 
\citep{nordlund+al_LRSP2009,muthsam+al2010}.
The behavior of vorticity in a gravitationally stratified atmosphere was
studied by \citet{arendt1993a,arendt1993b}. Since the gradients of
pressure and density are close to vertical, their cross product is 
horizontal, which explains the horizontal orientation of the
generated vorticity according to Equation~(\ref{eq:vorticity}).
This generation is different from the generation of 
vortical flows by conservation of angular momentum. The latter
mechanism is probably responsible for the vortical flows discovered
by \citet{bonet+al2008,bonet+al2010_shortt}.

Vortex tubes are a fundamental structure element of turbulence 
\citep{frisch1995,davidson2004}, where they play a key role in transferring
energy from large scales to the smallest, dissipative scales in the
energy cascade. It is tempting to speculate that the vortex tubes discovered in
the present investigation may represent just one stage in a similar energy 
cascade. They could potentially transport mechanical energy to different locations
and scales. The size of the vortex tubes seen in this investigation is close to the
spatial resolution limits of both the observation and the numerical simulation.
It should therefore be rewarding to further push these limits to smaller scales, as 
the Sun seems to provide a unique opportunity for the study of vortex tubes in a 
profoundly stratified medium.

\begin{acknowledgements}
The authors thank M.~Sch\"ussler for insightful discussions and the anonymous
referee for constructive comments.
The German contribution to {\sc Sunrise} is funded by the Bundesministerium
f\"{u}r Wirtschaft und Technologie through Deutsches Zentrum f\"{u}r Luft-
und Raumfahrt e.V. (DLR), Grant No. 50~OU~0401, and by the Innovationsfond of
the President of the Max Planck Society (MPG). The Spanish contribution has
been funded by the Spanish MICINN under projects ESP2006-13030-C06 and
AYA2009-14105-C06 (including European FEDER funds). The HAO contribution was
partly funded through NASA grant NNX08AH38G. The National Center for 
Atmospheric Research (NCAR) is sponsored by the National Science Foundation.
This work was supported by the WCU grant R31-10016 funded by the Korean 
Ministry of Education, Science, and Technology. 
\end{acknowledgements}


\bibliography{steiner+al}

\end{document}